\documentclass[twocolumn]{aastex63}
\usepackage{booktabs}
\usepackage{longtable}
\usepackage{xspace}
\usepackage{hyperref}
\usepackage[T1]{fontenc}
\usepackage{lipsum}
\usepackage{xstring} 

\newcommand{\Kepler}{\textit{Kepler}\xspace} 
\newcommand{\Gaia}{\textit{Gaia}\xspace}

\newcommand{\Mstar}{\ensuremath{M_{\star}}\xspace}
\newcommand{\Rstar}{\ensuremath{R_{\star}}\xspace}

\newcommand{\teff}{\ensuremath{T_{\mathrm{eff}}}\xspace}

\newcommand{\Rp}{\ensuremath{R_p}\xspace}

\newcommand{\rhostar}{\ensuremath{\rho_{\star}}\xspace}
\newcommand{\rtau}{\ensuremath{R_{\tau}}\xspace}

\newcommand{\sqrtecosw}{\ensuremath{\sqrt{e} \cos \omega}\xspace}
\newcommand{\sqrtesinw}{\ensuremath{\sqrt{e} \sin \omega}\xspace}

\newcommand{\Me}{\ensuremath{M_{\oplus}}\xspace} 
\renewcommand{\Re}{\ensuremath{R_{\oplus}}\xspace}

\newcommand{\Msun}{\ensuremath{M_{\odot}}\xspace}

\newcommand{\K}{\ensuremath{\mathrm{K}}\xspace}

\newcommand{\RNum}[1]{\uppercase\expandafter{\romannumeral #1\relax}}

\newcommand{\hand}[1]{
\IfEqCase{#1}{%
{n-f18}{907}%
{n-v18}{117}%
{n-f18-v18}{88}%
{prad-f18-v18-rms}{6.9\%}%
{srad-f18-v18-mean}{1.6\% larger}%
{srad-f18-v18-rms}{2.7\%}%
{ror-f18-v18-mean}{1.4\% smaller}%
{ror-f18-v18-rms}{6.6\%}%
{ror-t18-m15-rms}{6.8\%}%
{ror-dr25-v18-rms}{4.8\%}%
{ror-m15-ferr-mean}{3.1\%}%
{ror-t18-ferr-mean}{5.1\%}%
{ror-dr25-v18-rtau-rms}{3.6\%}%
}[XX]%
}

\begin{document}
\pagenumbering{arabic}



\title{Two Views of the Radius Gap and the Role of Light Curve Fitting}

\author[0000-0003-0967-2893]{Erik A.\ Petigura}
\affiliation{Department of Physics \& Astronomy, University of California Los Angeles, Los Angeles, CA 90095, USA}

\begin{abstract}
Recently, several groups have resolved a gap that bifurcates planets between the size of Earth and Neptune into two populations. The location and depth of this feature is an important signature of the physical processes that form and sculpt planets. In particular, planets residing in the radius gap are valuable probes of these processes as they may be undergoing the final stages of envelope loss. Here, we discuss two views of the radius gap by Fulton \& Petigura (2018; F18) and Van Eylen et al. (2018; V18). In V18, the gap is wider and more devoid of planets. This is due, in part, to V18's more precise measurements of planet radius \Rp. Thanks to \Gaia, uncertainties in stellar radii \Rstar are no longer the limiting uncertainties in determining \Rp for the majority of \Kepler planets; instead, errors in \Rp/\Rstar dominate. V18's analysis incorporated short-cadence photometry along with constraints on mean stelar density that enabled more accurate determinations of \Rp/\Rstar. In the F18 analysis, less accurate \Rp/\Rstar blurs the boundary the radius gap. The differences in \Rp/\Rstar are largest at high impact parameter ($b \gtrsim 0.8$) and often exceed 10\%. This motivates excluding high-$b$ planets from demographic studies, but identifying such planets from long-cadence photometry alone is challenging. We show that transit duration can serve as an effective proxy, and we leverage this information to enhance the contrast between the super-Earth and sub-Neptune populations.
\end{abstract}


\section{Introduction}
\label{sec:intro}

\begin{figure*}
\centering

\includegraphics[width=0.45\textwidth]{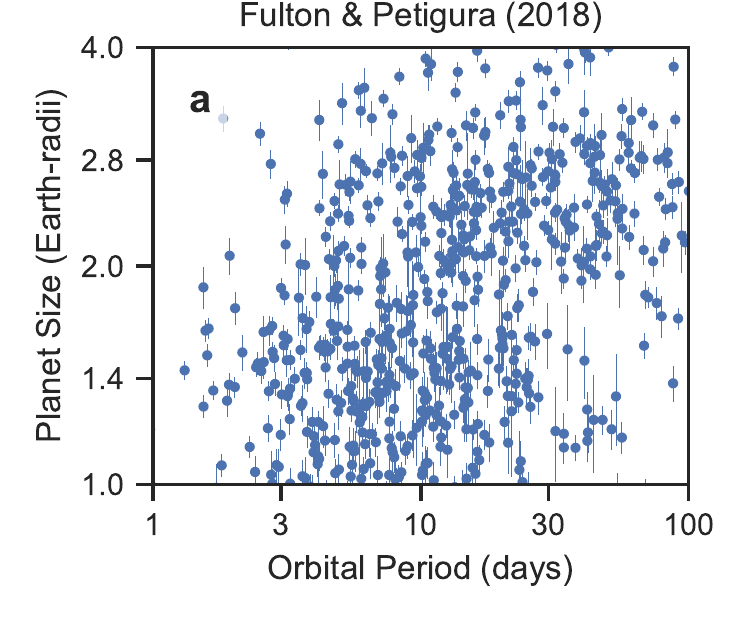}
\includegraphics[width=0.45\textwidth]{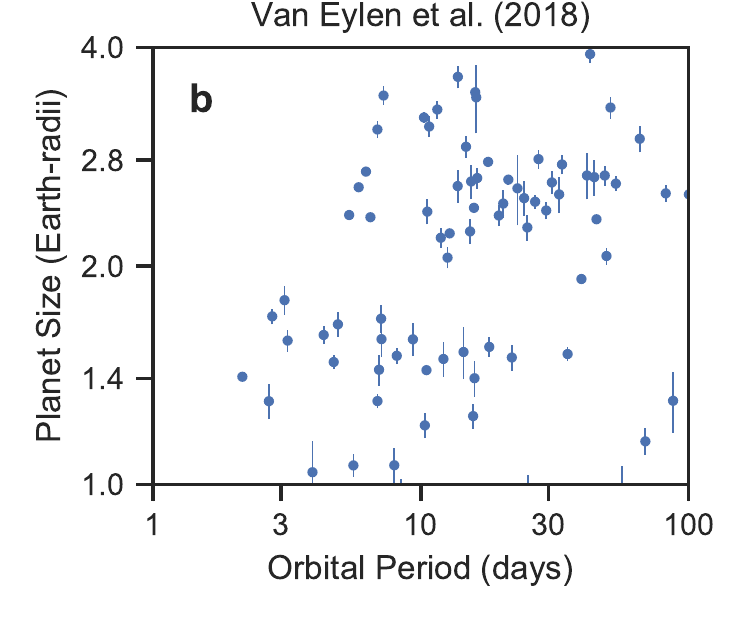}
\includegraphics[width=0.45\textwidth]{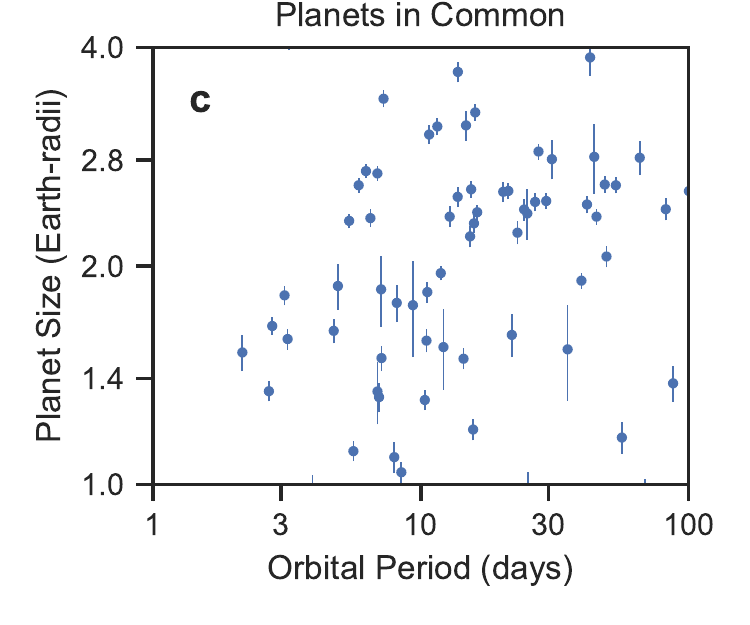}
\includegraphics[width=0.45\textwidth]{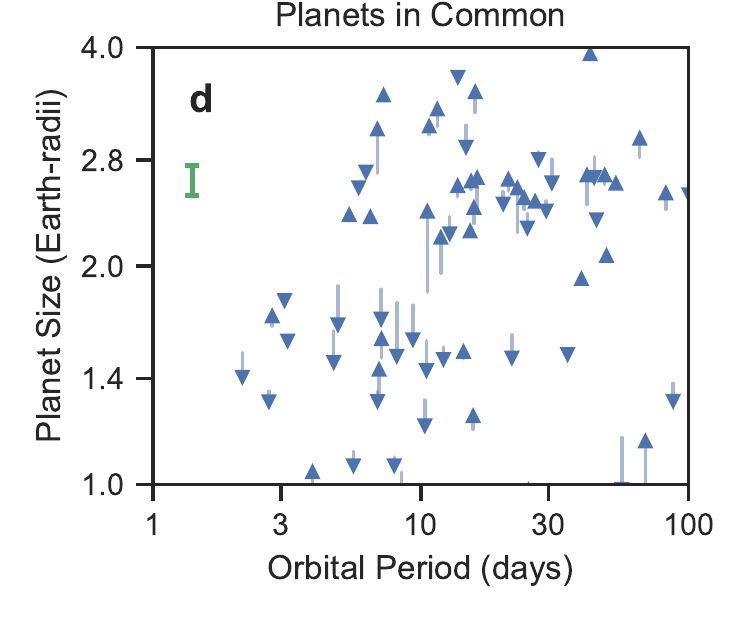}
\caption{Panel (a): sizes and orbital periods of planets from Fulton \& Petigura (2018; F18). Panel (b): same as (a) but from Van Eylen et al. (2018; V18). The radius gap is an underdensity of planets with \Rp = 1.5--2.0~\Re and is more devoid of planets in V18. Panel (c) just shows the F18 parameters for \hand{n-f18-v18} planets in common, but the gap is still less distinct. Panel (d): same as (c), but arrows indicate the change when we replace F18 \Rp with V18 \Rp; notably, planets move out of the radius gap. Many of the planets filling in the gap have radii that differ by $\gtrsim$10\% (green bar). \label{fig:f18-v18}}
\end{figure*}

\begin{figure*}
\centering
\includegraphics[width=4.0in,trim={4.25cm 0.5cm 4.5cm 0.25cm}]{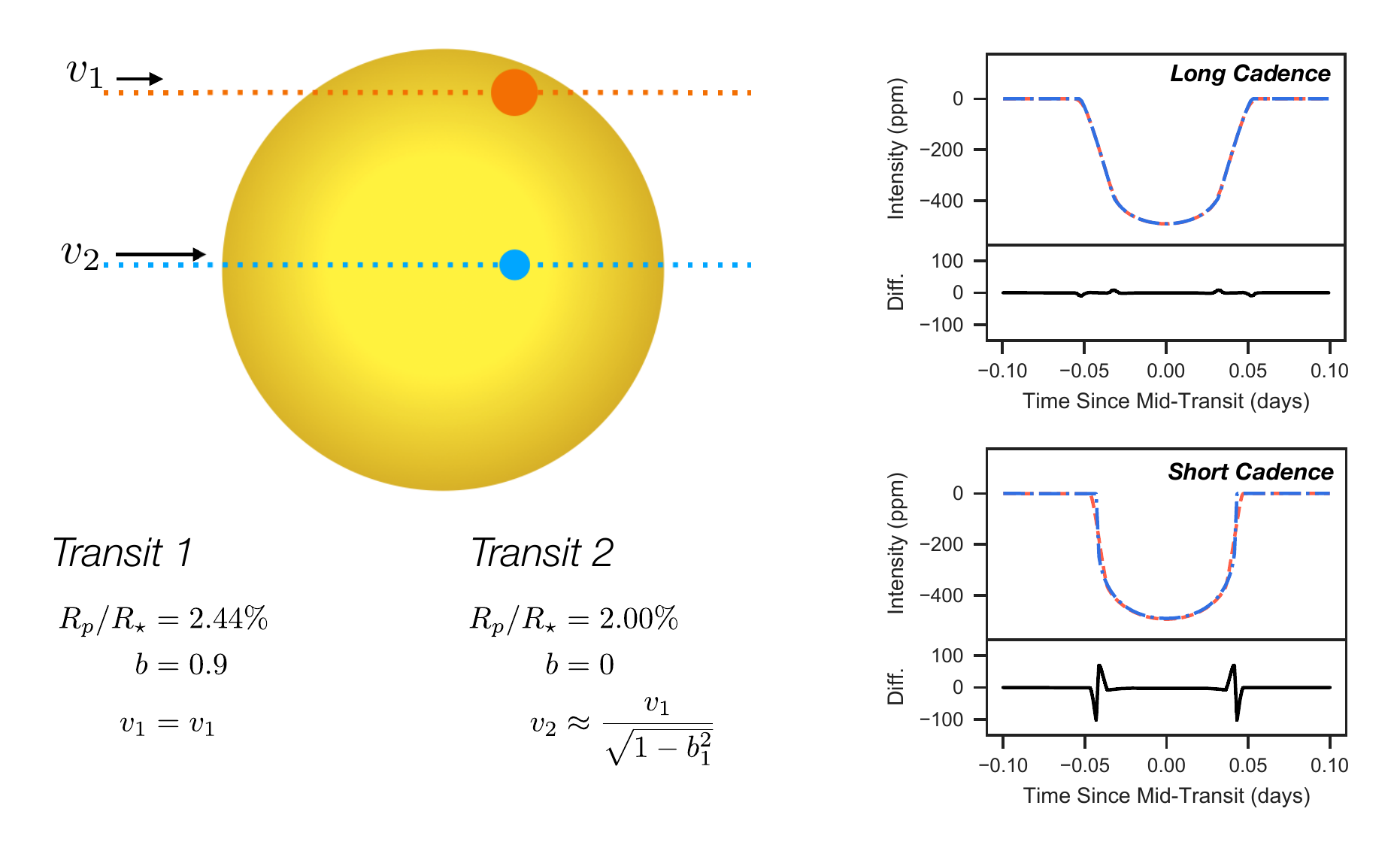}
\caption{Schematic illustration of \Rp/\Rstar errors that stem from poorly constrained impact parameter $b$. Transit~1 has a large, but non-grazing, $b$ and a size typical of \Kepler planets near the radius gap: $\left(\Rp/\Rstar\right)_1 = 2.44\%$, $b_1 = 0.9$. At top right, we show its long-cadence transit profile in orange. With Transit~2 (blue), we almost perfectly reproduced Transit~1 while requiring $b_2 = 0$ by (1) increasing the speed of the transit to account for the longer transit chord according to $v_2 \approx v_{1} / \sqrt{1 - b_1^2}$ and (2) reducing $\left(\Rp/\Rstar\right)_2$ to 2.00\% to compensate for the increased stellar brightness underneath the transit chord. The residual differences in the two long-cadence light curves are 10~ppm or less. Note the 20\% fractional difference in \Rp/\Rstar for Transits~1 \& 2. At lower right, we show analogous short-cadence transits that differ by as much as 100~ppm. Short-cadence photometry of sufficient quality can discriminate between the two scenarios. In the absence of such data, prior knowledge of mean stellar density and eccentricity act as a prior on $v$, which provides additional leverage on $b$ and $\Rp/R_\star$. Here, limb-darkening $\{ u_1, u_2 \} = \{0.45, 0.23\}$, $P = 11.5$~days, $\rhostar = 1~\mathrm{g~cm}^{-3}$, and $e = 0$.}
\label{fig:diagram}
\end{figure*}

If a planet transits, one may measure the planet-to-star radius ratio \Rp/\Rstar by modeling the transit light curve, and if \Rstar is known, one may compute \Rp through simple multiplication. Until recently, the radii of the vast majority of \Kepler planet hosts were constrained by photometry alone \citep{Brown11}. With 40\% uncertainties, these photometric \Rstar were the dominant uncertainty in \Rp. Over time, numerous groups worked to improve these radii through spectroscopic and asteroseismic techniques, but early studies were typically limited to a few hundred stars (\citealt{Furlan17}).

The California-\Kepler Survey (CKS) obtained high-resolution spectra of 1305 \Kepler stars, enabling \Rstar measurements good to 10\% \citep{Petigura17a,Johnson17}. Using these refined parameters, \cite{Fulton17} resolved a gap in the radius distribution of small planets spanning \Rp = 1.5--2.0~\Re. This gap was predicted qualitatively by several groups who considered the effects of XUV-driven photoevaporation of H/He envelopes surrounding cores of several \Me \citep{Owen13,Lopez13,Jin14,Chen16}. Since its discovery, the radius gap has spawned a flurry of theoretical interpretations that include star-powered and core-powered mass-loss mechanisms (e.g., \citealt{Owen17} and \citealt{Gupta19}). In a follow-up paper, \cite{Fulton18b}, F18 hereafter, further refined \Rstar by incorporating \Gaia parallaxes, 2MASS photometry, and spectroscopic \teff. F18 measured \Rstar with $3\%$ uncertainties and reported 5\% median uncertainties in \Rp.
 
Independent of CKS efforts, \cite{Van-Eylen18a}, V18 hereafter, observed the  gap in a smaller sample of \hand{n-v18} planets orbiting stars with asteroseismic detections. V18 achieved a median precision in \Rp of 3\% by modeling short-cadence photometry with asteroseismic constraints on \rhostar and \Rstar.

Figure~\ref{fig:f18-v18} compares the F18 and V18 views of the radius gap, which is wider and more devoid of planets in V18. F18 found that the location of the gap in \Rp grows as stellar mass \Mstar increases and provides a partial explanation: most of the V18 hosts are more massive than the sun, while F18 spans a larger range of $\Mstar \approx 0.8$--1.4~\Msun.   Differences in \Mstar, however, are not a complete explanation. Figure~\ref{fig:f18-v18} shows the F18 parameters for the \hand{n-f18-v18} planets common to both studies; the gap is still less distinct.

In this paper, we demonstrate that differences in \Rp/\Rstar also contribute to these different views. The differences are largest at high impact parameter $b$ due to differences in short- and long-cadence photometry and details in the fitting methodology (\S\ref{sec:comparison}). Figure~\ref{fig:diagram} summarizes this effect. For most \Kepler long-cadence transits, there are large uncertainties in $b$. When the differences between the inferred and true $b$ are large, we find large discrepancies between the modeled and true \Rp/\Rstar due to the effects of limb-darkening. While these effects have been studied analytically in previous works \citep{Carter08,Price14}, we focus specifically on the CKS sample and the radius gap. In \S\ref{sec:filter}, we use MCMC techniques and transit duration information to improve the agreement between F18 and V18 by a factor of two. Applying these techniques to the CKS sample enhances the contrast between the super-Earth and sub-Neptune populations (\S\ref{sec:cks}). We also offer some suggestions for further refinements of \Kepler planet radii (\S\ref{sec:future-work}).

\section{Comparing F18 and V18}
\label{sec:comparison}

\subsection{\Rp/\Rstar Dominates \Rp Dispersion}
\label{sec:radius-ratio}

\begin{figure*}[h]
\centering
\hspace{-0.5cm}
\includegraphics[width=0.34\textwidth]{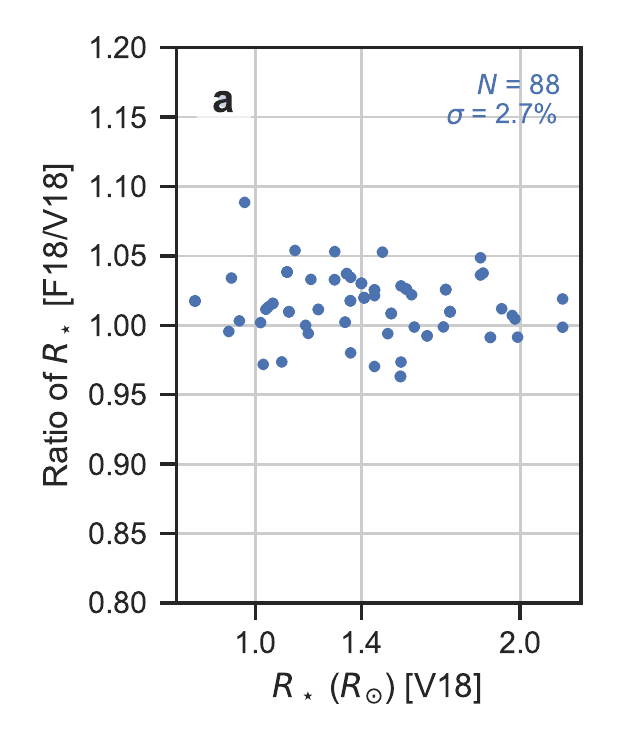}
\hspace{-0.5cm}
\includegraphics[width=0.34\textwidth]{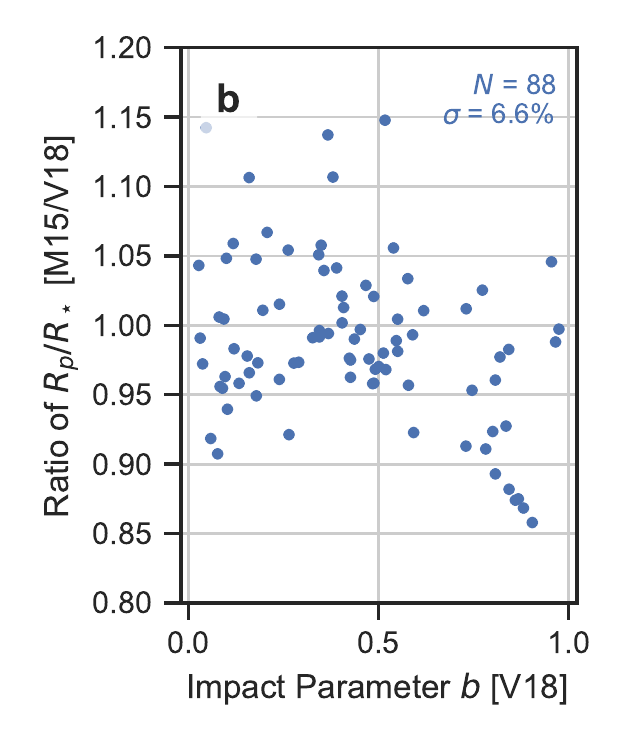}
\hspace{-0.5cm}
\includegraphics[width=0.34\textwidth]{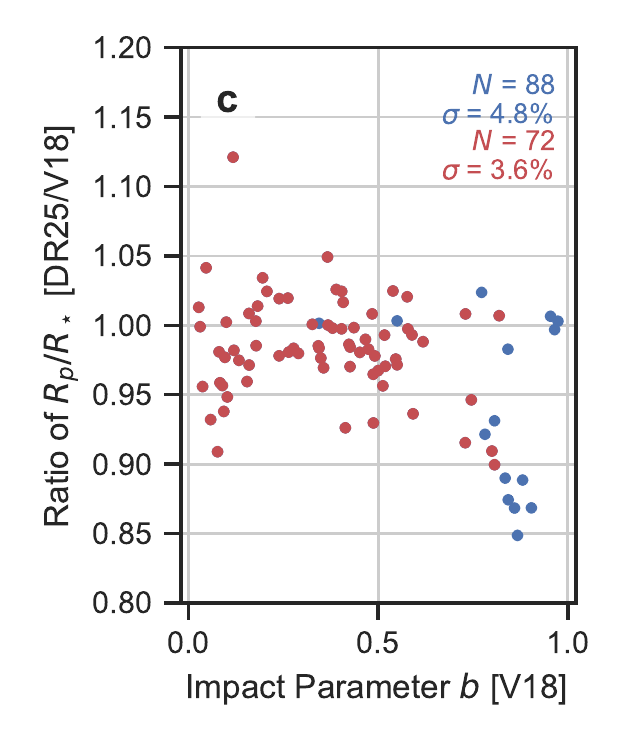}
\caption{Comparison of \Rstar and \Rp/\Rstar from different catalogs. Panel (a): ratio of \Rstar from F18 and V18. The \hand{srad-f18-v18-rms} dispersion in \Rstar cannot explain the \hand{prad-f18-v18-rms} dispersion in \Rp. Panel (b):  ratio of \Rp/\Rstar from Mullally et al. (2015; M15) and V18. The \hand{ror-f18-v18-rms} dispersion accounts for the majority of the \Rp dispersion (\S\ref{sec:comparison}). Panel (c): After swapping the M15 best-fit \Rp/\Rstar with the DR25 posterior medians, the dispersion decreases to \hand{ror-dr25-v18-rms}.  After applying a transit duration filter designed to remove planets with $b \gtrsim 0.8$, the remaining planets (red points) have a \hand{ror-dr25-v18-rtau-rms} dispersion in \Rp/\Rstar (\S\ref{sec:filter}).
\label{fig:compare}}
\end{figure*}

We begin by comparing the \Rp derived by F18 and V18 for \hand{n-f18-v18} planets in common. Among these planets, there is a \hand{prad-f18-v18-rms} RMS dispersion in the ratio of \Rp. Because \Rp depends on both \Rp/\Rstar and \Rstar, we compare these two quantities separately in  Figure~\ref{fig:compare}. There is a \hand{srad-f18-v18-rms} dispersion in \Rstar that is insufficient to account for the majority of the \Rp dispersion. Instead, the majority of the \Rp dispersion stems from differences in \Rp/\Rstar, which have a dispersion of \hand{ror-f18-v18-rms}. F18 used $\Rp/\Rstar$ computed in \cite{Mullally15}, M15 hereafter, while V18 derived \Rp/\Rstar independently. We compare these two approaches in detail in \S\ref{sec:noise}.

%

\subsection{The Role of Impact Parameter}
\begin{figure*}
\centering
\includegraphics[width=0.49\textwidth]{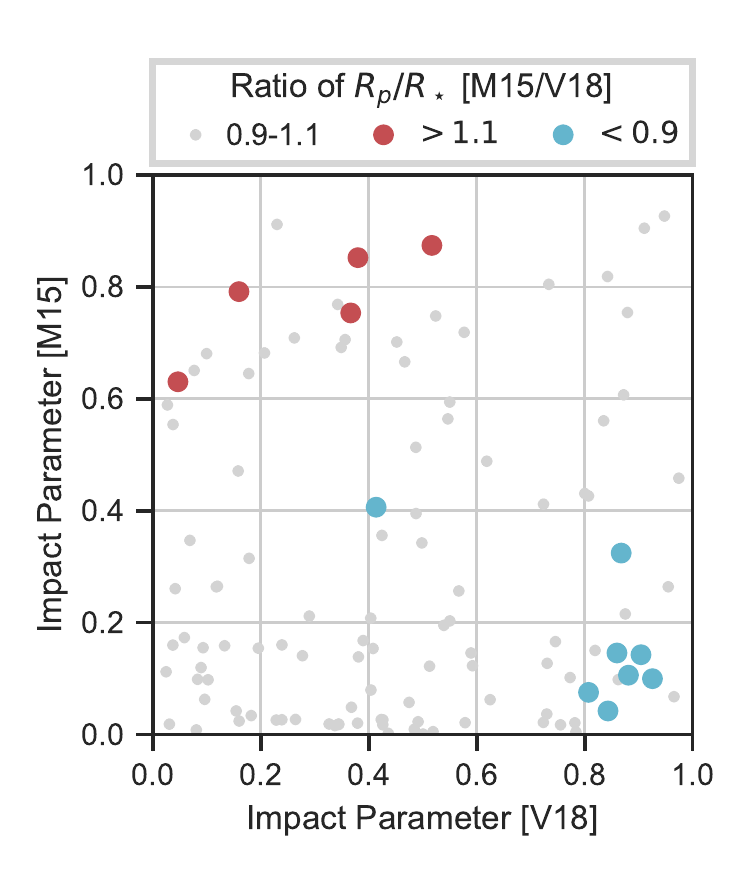}
\includegraphics[width=0.49\textwidth]{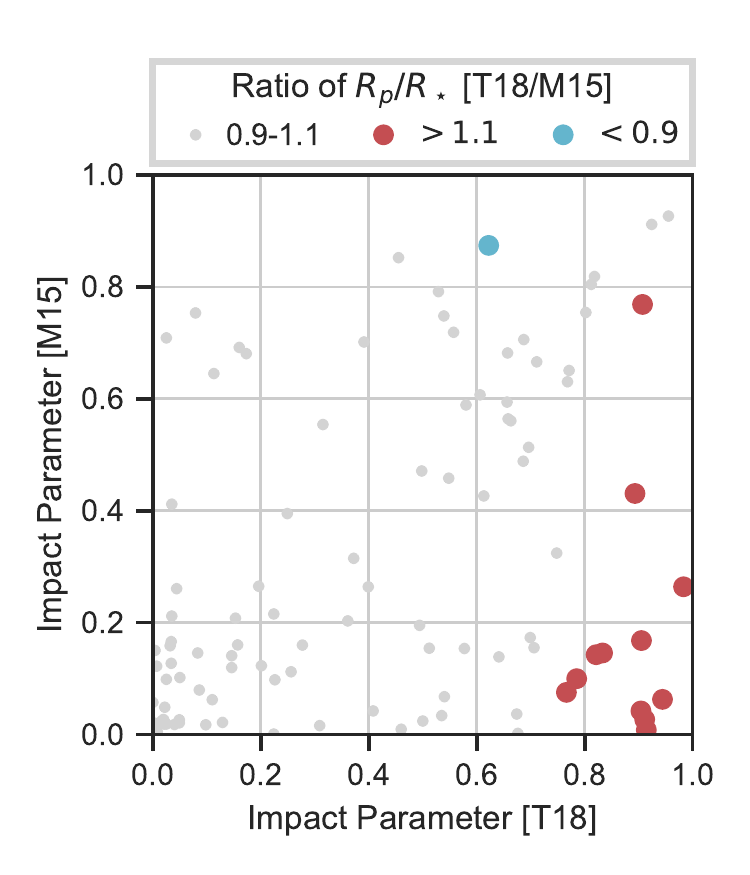}
\caption{Left: comparison of impact parameters derived by V18 and M15 showing little correlation between the two studies. Planets with the most discrepant \Rp/\Rstar are shown as large points and occur when one analysis favors high $b$, while the other does not. Right: same as left, but comparing M15 and T18. The numerical instabilities inherent to the best-fit values are apparent as both studies analyzed nearly slightly different reductions of the same photometry, yet sometimes returned \Rp/\Rstar that differed by more than 10\%.  \label{fig:impact-v18-m15}}
\end{figure*}

Comparing the impact parameters from M15 and V18 provided a clue to understanding the discrepant \Rp/\Rstar. Figure~\ref{fig:impact-v18-m15} shows $b$ from both analyses, which are remarkably uncorrelated. The most discrepant \Rp/\Rstar tend to occur when either M15 or V18 favored high $b$, while the other did not. Because stellar surface brightness decreases toward the limb, the fit favoring high $b$ must increase \Rp/\Rstar to match the transit depth.

\subsection{Noise from Fitting}
\label{sec:noise}
Some of the dispersion between M15 and V18 \Rp/\Rstar is due to details in the fitting methodology, which we summarize below. M15 fit transits using the \cite{mandel:2002} model, which includes the following parameters: $\{P, T_c, \Rp/\Rstar, a/\Rstar, b, u_1, u_2, F\}$. Here, $u_1$ and $u_2$ are quadratic limb-darkening coefficients and $F$ is an overall flux normalization. Instead of fitting in terms of $a/\Rstar$, M15 fit 
\begin{equation}
\label{eqn:rhostartr}
\rho_{\star,tr} = \frac{3 \pi}{G P^2} \left( \frac{a}{\Rstar}\right)^3,
\end{equation}
where the subscript serves to distinguish this reparameterization of $P$ and $a/\Rstar$ from the star's true density. The M15 parameters on the NASA Exoplanet Archive (NEA; \citealt{Akeson13}) are best-fit values as determined by a Levenberg-Marquardt algorithm and the uncertainties are determined by MCMC.%
%

V18 used the same Mandel-Agol description. V18 first fit for $P$ after removing any TTVs, and fixed it in their subsequent MCMC analysis. Instead of sampling directly in $a/\Rstar$, V18 sampled in $\{\sqrtecosw,\sqrtesinw\}$ and used Equation~\ref{eqn:rhostartr} and the following equation to calculate $a/\Rstar$:
\begin{equation}
\frac{\rhostar}{\rho_{\star,tr}} = \frac{\left(1 - e^2\right)^{3/2}}{\left(1 + e \sin \omega \right)^3}.
\end{equation}
V18 fixed \rhostar  to the asteroseismic value. An important distinction between the two methodologies is V18 enforced a uniform prior on $e$ that takes into account \rhostar. In contrast, M15 did not incorporate \rhostar, and sampling $\rho_{\star,tr}$ implicitly introduces a non-uniform $e$ prior \citep{Dawson12}.

We found that the Levenberg-Marquardt fitting scheme used by M15 is numerically unstable and introduced noise into $\Rp/\Rstar$. This can be seen by comparing $\Rp/\Rstar$ measured by M15 and \cite{Thompson18}, T18 hereafter. T18 used the same fitting procedure for a slightly different reduction of the same photometry.%
\footnote{While M15 identified candidates in Q1--Q16 photometry, Q17 was included in the final modeling. T18 also modeled Q1--Q17. M15 used DR21--DR23, while T18 used DR25.}
For the planets in our comparison sample, there is a \hand{ror-t18-m15-rms} dispersion between the M15 and T18 \Rp/\Rstar, larger than their average reported uncertainties of \hand{ror-m15-ferr-mean} and \hand{ror-t18-ferr-mean}, respectively. Given the similarity of the input photometry, we should expect the dispersion in \Rp/\Rstar to be {\em smaller} than the formal uncertainties. As we show below, $b$ is often nearly unconstrained from long-cadence photometry and variability in the best-fit $b$ is associated with variability in $\Rp/\Rstar$.

\begin{figure*}
\centering
\includegraphics[width=0.49\textwidth]{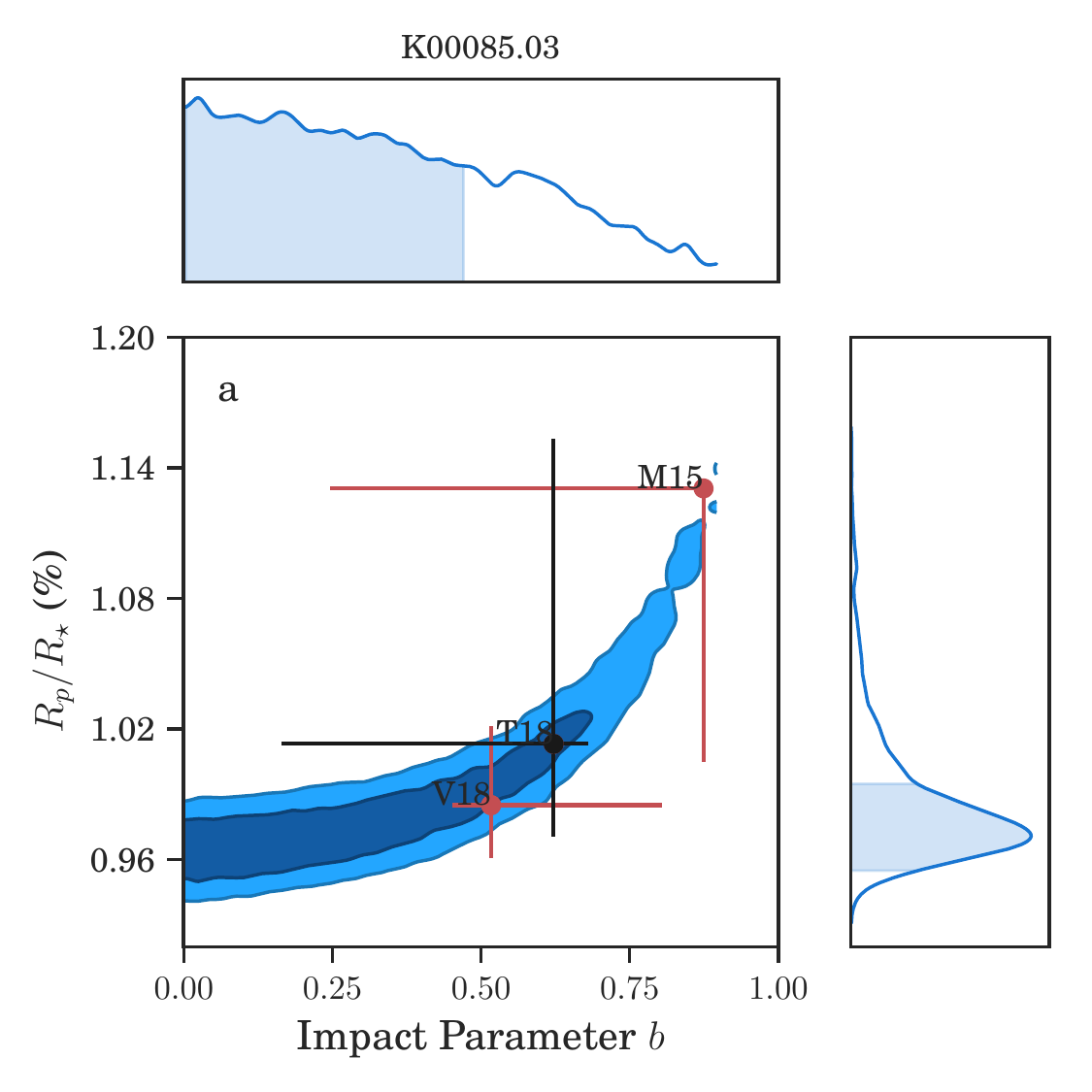}
\includegraphics[width=0.49\textwidth]{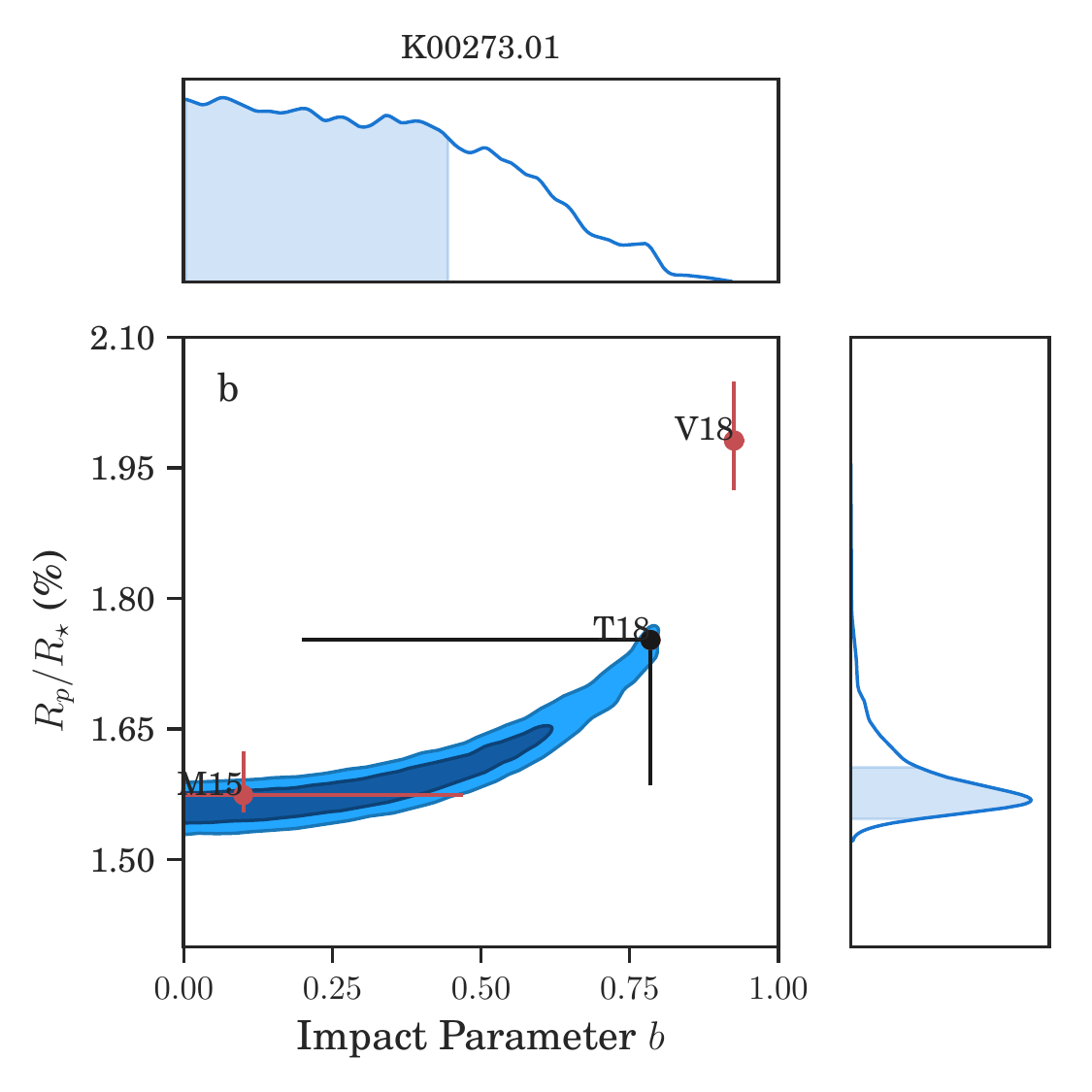}
\caption{Two planets residing in the F18 gap that illustrate two failure modes of long-cadence fitting. Panel (a) shows different reported radius ratios and impact parameters for K00085.03. Here, the M15 least-squares fitter terminated at high $b$ and high $\Rp/\Rstar$, while the V18 analysis favored a smaller $b$ and $\Rp/\Rstar$. The contours are the 1$\sigma$ and 2$\sigma$ levels from T18's MCMC analysis of long-cadence photometry. Panel (b): same as (a), but for K00273.01. In this case, the V18 fits favored high $b$, while the M15 fit terminated at low $b$.\label{fig:corner}}
\end{figure*}

Figure~\ref{fig:corner} shows the joint $\Rp/\Rstar$-$b$ posteriors from the T18 analysis for two planets near the radius gap. For both planets, $b$ is nearly unconstrained and is highly correlated with $\Rp/\Rstar$. For K00085.03, the bulk of the posterior is between $b$ = 0.0--0.8, but the M15 fitter terminated at $b = 0.93$. Perhaps the optimizer got stuck in a local minimum or failed to trace the curving $\chi^2$ surface. For the same planet, V18 favored lower $b$ and lower \Rp/\Rstar. In contrast, for K00273.01, the M15 fitter terminated at low $b$, but the V18 analysis detected a large $b$ with high significance and returned a larger $\Rp/\Rstar$.

Between the M15, T18, and V18 catalogs of \Rp/\Rstar, we suspect that V18 is the closest to the ground truth because it incorporated short-cadence photometry, imposed priors on \rhostar, and reported the posterior medians. The dispersion in our M15-V18 and M15-T18 comparisons point toward additional noise in M15 \Rp/\Rstar from the fitting, which introduced additional uncertainty into the F18 \Rp. In the following section, we recommend two modifications to the F18 \Rp to improve precision: (1) adopt the posterior medians, which do not suffer from the numerical noise of the best-fit values, and (2) discard high-$b$ transits where \Rp/\Rstar is often biased away from the true value.

\section{Improving the Radius Ratios}
\label{sec:filter}

Figure~\ref{fig:corner}a illustrates a danger of adopting best-fit parameters, which can favor high $b$ even when it is not needed to fit the data. We recommend using the DR25 posterior medians, which are more robust estimators of central tendancy. Figure~\ref{fig:compare}c shows the ratio of the posterior medians of $\Rp/\Rstar$ from DR25 compared to V18. The RMS dispersion has decreased from \hand{ror-f18-v18-rms} to \hand{ror-dr25-v18-rms}. 

The majority of the remaining outliers in Figure~\ref{fig:compare}b are planets where V18 measured $b > 0.8$. K00273.01 is one such outlier. Here, the long-cadence sampling could not resolve the high $b$ and the bulk of the posterior resides below $b$ of 0.8. 

For typical \Kepler planets, it is not possible constrain $b$ sufficiently with long-cadence photometry to reliably discard high-$b$ planets. However, one may use transit duration as a proxy for $b$ when \rhostar is known. When $\Rp \ll \Rstar$, the time between ingress and egress midpoints is given in Equation~14 of \cite{Winn10d}:

\begin{equation}
\label{eqn:tau-full}
\tau = \frac{P}{\pi}
           \sin^{-1} 
           \left( 
               \frac{\Rstar}{a}
               \frac{\sqrt{1- b^2}}{\sin i}
           \right) 
       \frac{\sqrt{1-e^2}}{1+e \sin{\omega}}.
\end{equation}
We define $\tau_0$ as $\tau$ when $b = 0$ and $e = 0$. Plugging in Equation~\ref{eqn:rhostartr}, 
\begin{eqnarray}
\tau_{0} & = &  
   \frac{P}{\pi}
           \sin^{-1} 
           \left[ 
           \left(
               \frac{3 \pi}{G P^2 \rhostar}
           \right)^{1/3}
           \right].\label{eqn:tau0-rho}
\end{eqnarray}
We define the ratio between these two durations as
\begin{equation}
\rtau = \tau / \tau_0.
\end{equation}
For transiting planets, $\Rstar/a \ll 1$  and $\sin i \approx 1$ so, 
\begin{equation}
\label{eqn:rtau-approx}
\rtau = \sqrt{1 - b^2} \frac{\sqrt{1-e^2}}{1+e \sin{\omega}} + \mathcal{O}\left(\frac{\Rstar}{a}\right)^2 .
\end{equation}
If \rhostar is known exactly, $\rtau < 1$ implies either $b > 0$ or $e \sin \omega > 0$; if $\rtau > 1$, $e \sin \omega$ must be negative. Therefore, one may use \rtau to identify high-$b$ transits, with eccentricity as a complicating factor. Looking ahead to \S\ref{sec:cks}, we will compute $\tau$ from MCMC modeling of \Kepler long-cadence photometry and compute $\tau_{0}$ from $P$, $\rhostar$, and Equation~\ref{eqn:tau0-rho}. We will then filter out high-$b$ transits from the CKS sample. 

Before proceeding to the CKS dataset, we investigated the complicating effects of eccentric orbits and measurement uncertainties on such a filter. We simulated five transit surveys, described below.

\begin{figure*}
\centering
\includegraphics[width=0.32\textwidth]{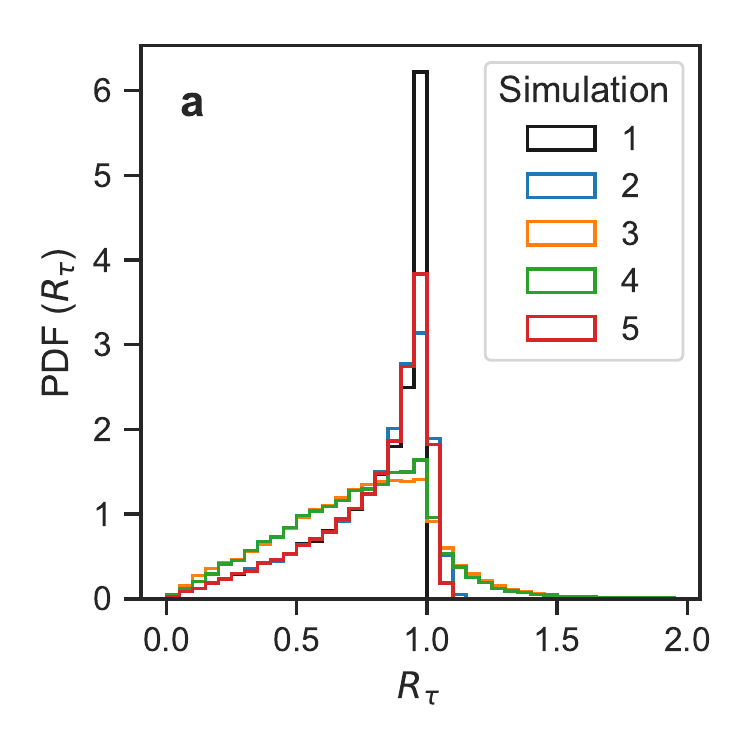}
\includegraphics[width=0.32\textwidth]{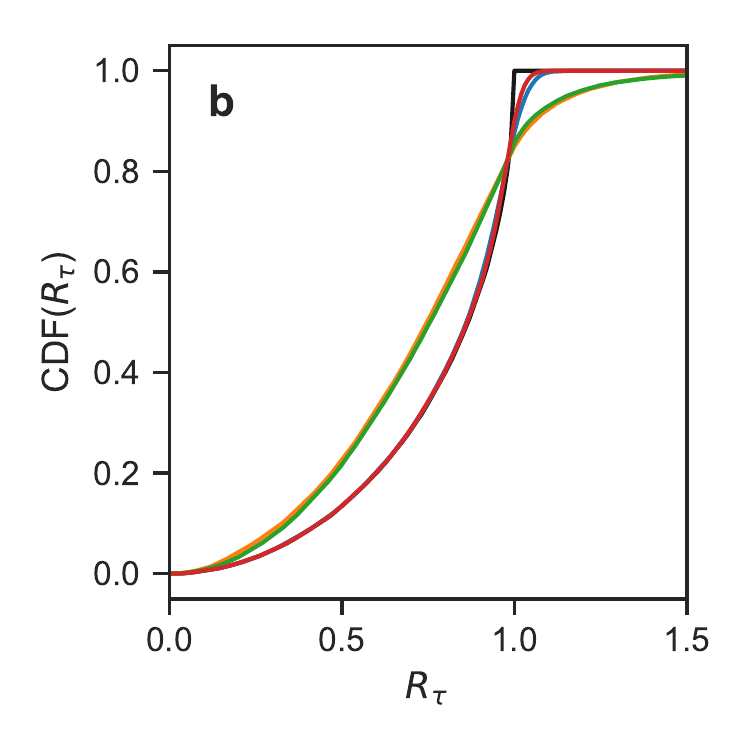}
\includegraphics[width=0.32\textwidth]{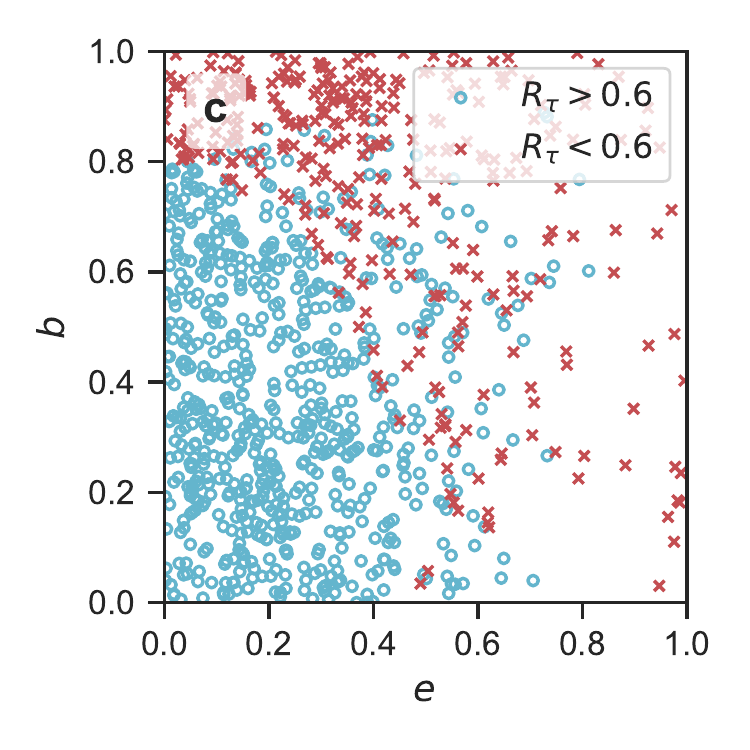}
\caption{We computed \rtau, the ratio of the observed transit duration to the maximum duration for a planet on a circular orbit, for five different simulated planet populations described in Table~\ref{tab:simulations}. 
Panel (a) shows the probability density function (PDF) and panel (b) shows the cumulative  distribution function (CDF). Panel (c) shows eccentricity $e$ and impact parameter $b$ from a random sampling of 1,000 planets from Simulation~3. Planets where $\rtau > 0.6$ or $\rtau < 0.6$ are indicated with different colors/symbols. Removing planets with $\rtau < 0.6$ discards 90\% of the planets with $b$ > 0.8, resulting in a filtered sample where 97\% of the planets have $b$ < 0.8.\label{fig:tau-hist}}
\end{figure*}

\begin{enumerate}
\item {\em Baseline.} To build intuition, we first simulated a population of $10^6$ planets by drawing $\cos{i}$ from a uniform distribution and setting $e = 0$. To compute \rtau, we arbitrarily set $P = 20$~days and $\rhostar$ = 1.4 g/cc. However, Equation~\ref{eqn:rtau-approx} shows that $P$ and \rhostar lead to corrections in \rtau that are second order in $\Rstar/a \ll 1$. We also assumed that \rhostar and $\tau$ are measured exactly. The input parameters to all our simulations are summarized in Table~\ref{tab:simulations}. Figure~\ref{fig:tau-hist} displays the probability density function (PDF) and cumulative distribution function (CDF) of $\rtau$. Most of the transits have durations near $\tau_0$, but there is a tail to low \rtau from high $b$.

\item{\em Uncertain mean stellar density.} We then simulated uncertainties in \rhostar. We computed $\tau_0$ using $\rho_{\star,1}$ that differed from the input $\rhostar$. We drew $\log_{10}(\rho_{\star,1}/\rhostar)$ from a normal distribution $\mathcal{N}(0,0.06)$, i.e. 15\% fractional precision---typical for F18. In Figure~\ref{fig:tau-hist}, \rtau now spills past unity, but the overall change is minor.

\item{\em Eccentricity.} The eccentricity distribution is one of the key characteristics of the exoplanet population, and characterizing it as been the subject of many works (see \citealt{Winn15} for a review). Using the same dataset as V18, \cite{Van-Eylen19} measured the eccentricities of individual \Kepler planets through detailed modeling of short-cadence photometry and \rhostar from asteroseismology. They found that the eccentricities of singles and multis were well-described by positive (one-sided) Gaussians of $\mathcal{N}^{+}(0,0.3)$ and $\mathcal{N}^+(0,0.08)$, respectively. In Simulation 3, we drew $e$ from the broader $\mathcal{N}^{+}(0,0.3)$ distribution and $\omega$ from a uniform distribution. Compared to Simulations 1 and 2, the distribution of \rtau is more dispersed.

\item{\em Eccentricity---alternate distribution.} We consider the \cite{Van-Eylen19} results to be the state-of-the-art description of \Kepler planet eccentricities. However, our results above do not depend sensitively on that analysis. For example, \cite{Kipping13} characterized the eccentricity distribution of RV planets (mostly more massive than 100~\Me) with a two-parameter beta distribution $\mathcal{B}(0.867,3.03)$. The distribution of $\rtau$ is nearly identical to that of Simulation 3.

\item{\em Uncertainties in measured duration.} Finally, we considered the effect of uncertainties in $\tau$. For most planets, this is relatively minor because the fractional uncertainty is small. Among the T18 planet candidates (i.e. not dispositioned as ``False Positives'') the median fractional precision on this quantity is $3.5\%$. In Simulation 5, we simulated 3.5\% $\tau$ errors. The effect on the $\rtau$ distribution is comparable to the \rhostar effects considered in Simulation~1, but smaller than the eccentricity effects considered in Simulations~3 and 4.
 
\end{enumerate}

\begin{deluxetable}{lRRR}[h!]
\tablecaption{Simulated Transit Surveys\label{tab:simulations}}
\tabletypesize{}
\tablecolumns{4}
\tablehead{
	\colhead{Simulation} & 
	\colhead{$e$} & 
	\colhead{$\sigma(\rhostar)$} &
	\colhead{$\sigma(\tau)$} \\
	\colhead{} & 
	\colhead{} & 
	\colhead{\%} &
	\colhead{\%}
}
\startdata
1 & 0                        & 0  & 0   \\
2 & 0                        & 15 & 0   \\
3 & \mathcal{N}^{+}(0,0.3)   & 0  & 0   \\
4 & \mathcal{B}(0.867, 3.03) & 0  & 0   \\
5 & 0                        & 0  & 3.5 
\enddata
\tablecomments{Summary of the five simulated transit surveys described in \S\ref{sec:filter}. We considered different exoplanet eccentricity distributions: circular, one-sided Gaussian $\mathcal{N}^{+}$, and beta $\mathcal{B}$. We also simulated the effects of 15\% uncertainties in mean stellar density \rhostar, and 3.5\% uncertainties in measured transit duration $\tau$.}
\end{deluxetable}

To summarize, in the limiting case where $e=0$ and where \rhostar and $\tau$ are known exactly, \rtau may be tuned to  remove an arbitrary range of $b$. Non-zero eccentricities, uncertainties in \rhostar, and uncertainties in $\tau$ reduce the purity of such a filter. Our simulations showed that eccentricity is the dominant complication. 

We used Simulation 3 to motivate a filter designed to remove high-$b$ transits based on their short durations. In Figure~\ref{fig:tau-hist}, we illustrate the consequences of applying a $\rtau < 0.6$ filter designed to remove $b > 0.8$ planets. We show $b$ and $e$ of a random sampling of 1,000 planets from Simulation~3. At low $e$, $\rtau = 0.6$ effectively divides the planets at $b = 0.8$. As eccentricity rises, the low- and high-b planets mix. The criterion $\rtau < 0.6$ correctly identifies $18\%/20\% = 90\%$ of the planets with $b > 0.8$ (true-positive rate). Of the remaining stars, only $2\%/68\% = 3\%$ have $b > 0.8$ (false-negative rate). However, this additional purity comes at the cost of removing $14\%/80\% = 18\%$ of the planets with $b < 0.8$ (false-positive rate).

Figure~\ref{fig:compare}c includes this filter on the F18-V18 comparison sample. Requiring $\rtau > 0.6$ removes 16/\hand{n-f18-v18} comparison stars, 12 of which have $b > 0.8$, according to V18. Only four stars with $b < 0.8$ are removed; the false-positive rate is 5\% compared to 18\% in Simulation~3.

The eccentricity distribution Simulation~3, $\mathcal{N}^{+}(0,0.3)$ with $\langle e \rangle$ = 0.24, should be interpreted as a bounding case. \cite{Van-Eylen19} found that the eccentricity distribution of multis is narrower, $\mathcal{N}^{+}(0,0.08)$ where $\langle e \rangle$ = 0.06. This agrees with \cite{Mills19} who found  $\langle e \rangle$ = 0.21 and 0.05 for singles and multis, respectively, using the larger CKS sample. Since the F18-V18 comparison sample contains both singles and multis, it is not surprising that our filter resulted in a lower false-positive rate on real data than in Simulation~3.

After applying this cut, the F18-V18 dispersion in \Rp/\Rstar is \hand{ror-dr25-v18-rtau-rms}. Together, the two techniques outlined in this section reduced this dispersion from \hand{ror-f18-v18-rms} to \hand{ror-dr25-v18-rtau-rms}---nearly a factor of two---at the cost of decreasing the sample size by 18\%.

\section{Refining The CKS Sample}
\label{sec:cks}

\begin{figure*}
\centering
\includegraphics[width=0.85\textwidth]{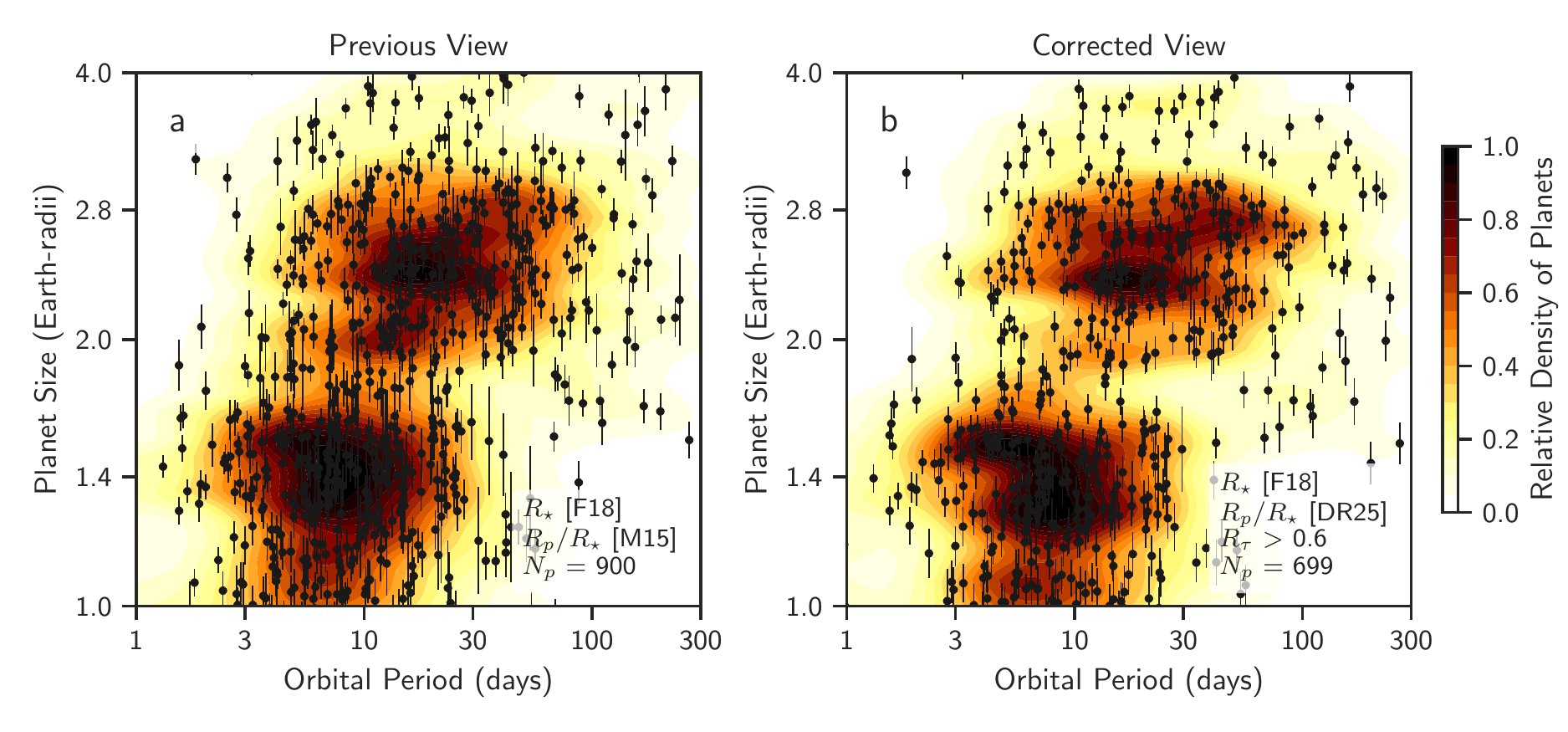}
\includegraphics[width=0.85\textwidth]{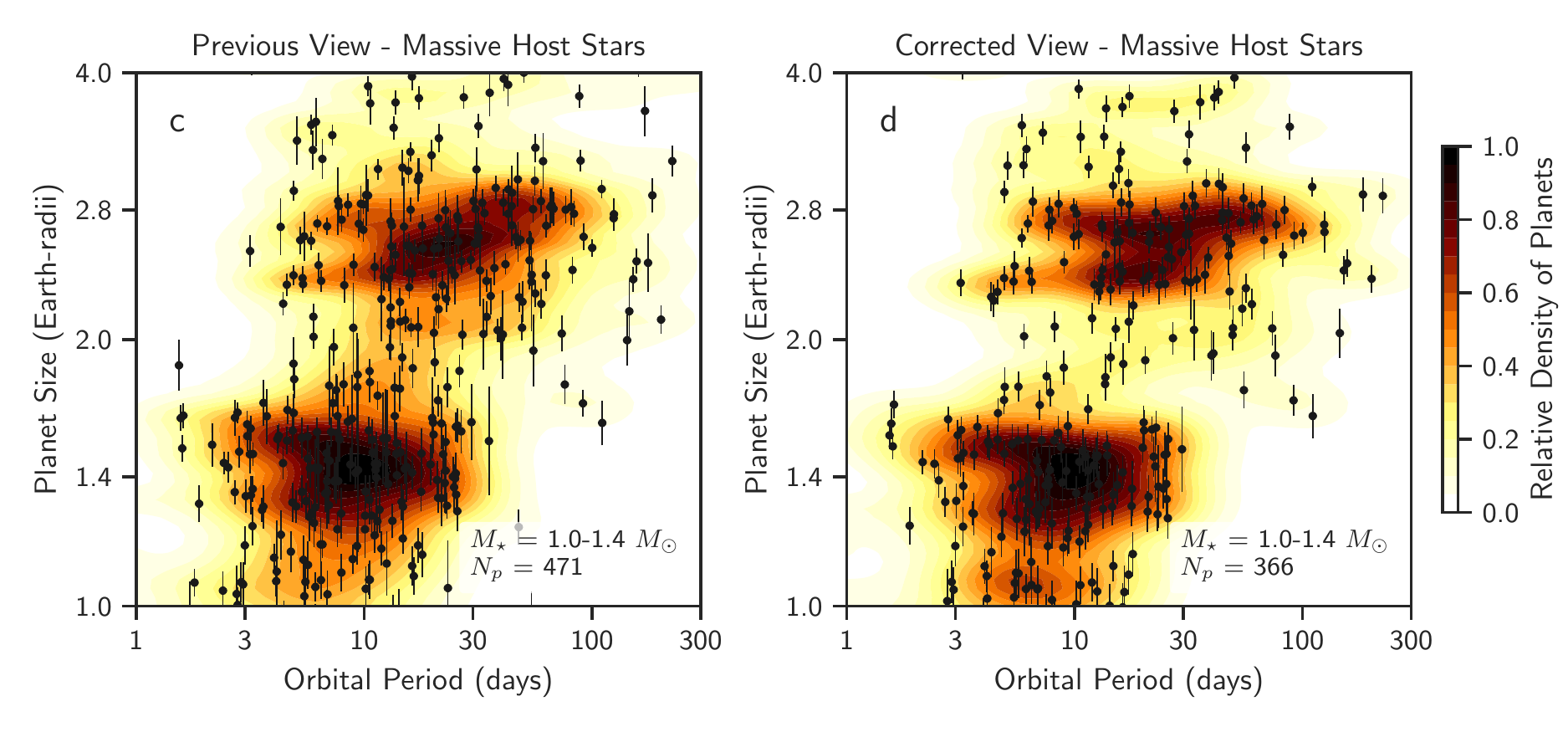}
\caption{The period-radius distribution of \Kepler planets illustrating the effects of different \Rp/\Rstar provenances and \Mstar distributions. Panel (a): The F18 sample where \Rp are based on F18 \Rstar and M15 \Rp/\Rstar. The super-Earth and sub-Neptune populations are resolved in this sample, but they are less distinct than in V18 (Figure~\ref{fig:compare}). Panel (b): same as (a) but we have replaced the best-fit M15 \Rp/\Rstar with posterior medians and have removed stars with anomalously short durations that are suggestive of $b \gtrsim 0.8$ (see \S\ref{sec:filter}). The two populations are more distinct in (b) than in (a). Some of the super-Earths with large positive fluctuations in $\Rp$ due to erroneously high $b$ in (a) have been adjusted down with the new \Rp/\Rstar; some of the sub-Neptunes with large negative fluctuations in \Rp due to undetected high $b$ have been removed by the \rtau cut. Panel (c): same as (a) with \Mstar = 1.0--1.4~\Msun.  Panel (d): same as (b) but for \Mstar = 1.0--1.4~\Msun. Of the four panels, panel (d) bears the closest resemblance to the V18 distribution shown in Figure~\ref{fig:compare}b.}
\label{fig:compare-per-prad}
\end{figure*}

Here, we apply the two techniques developed in the previous section to produce a new view of the radius gap. Figure~\ref{fig:compare-per-prad}a shows the F18 planet population in the $P$-\Rp plane. To construct Figure~\ref{fig:compare-per-prad}b, we replaced the M15 $\Rp/\Rstar$ with the DR25 posterior medians and removed planets with $\rtau < 0.6$. While the sample is $\approx$20\% smaller, the super-Earth and sub-Neptune populations are more distinct.

The increased contrast can be understood as follows: Some of the super-Earths in Figure~\ref{fig:compare-per-prad}a had erroneously high \Rp/\Rstar due to high M15 $b$ (e.g., K00085.03; Figure~\ref{fig:corner}a). The DR25 $\Rp/\Rstar$ posterior medians are lower and these planets join the bulk of the super-Earth population in Figure~\ref{fig:compare-per-prad}b. In addition, some of the sub-Neptunes had erroneously small \Rp/\Rstar because the long-cadence fits could not identify high $b$ (e.g., K00273.01; Figure~\ref{fig:corner}b). The duration filter removed most of these in Figure~\ref{fig:compare-per-prad}b.

In addition to \Rp precision, \Mstar also contributed to differences in the F18 and V18 views of the radius gap. Figure~\ref{fig:compare-per-prad}c shows the F18 parameters, but for $\Mstar$ = 1.0--1.4~\Msun, which approximates the V18 stellar mass distribution. Compared to the full sample, the super-Earth and sub-Neptune populations are more widely separated. Again, after swapping \Rp/\Rstar and applying the $\rtau$ cut in Figure~\ref{fig:compare-per-prad}d, we enhance the contrast between the two populations. 

Of all the samples shown in Figure~\ref{fig:compare-per-prad}, panel d most closely resembles the distribution in V18 in terms of contrast and separation between the two populations. One difference is the gap does not clearly have a negative slope. We note that the negative slope in V18 is driven by just a few planets and that quantifying the slope of the absence of planets is a challenging statistical problem. 

\begin{figure}
\includegraphics[width=0.49\textwidth]{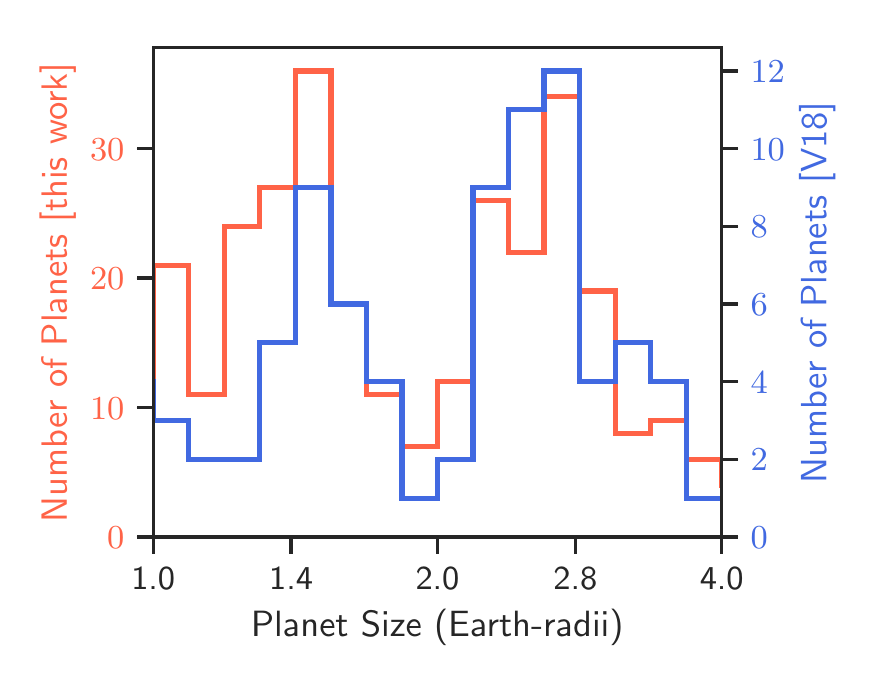}
\caption{A comparison of the planet radius distributions from this work (red; Figure~\ref{fig:compare-per-prad}d) and V18 (blue; Figure~\ref{fig:f18-v18}b).}
\label{fig:compare-v18-thiswork-hist}
\end{figure}

We provide a final comparison between our corrected view of the high-\Mstar sample (Figure~\ref{fig:compare-per-prad}d) and the V18 sample (Figure~\ref{fig:f18-v18}b) in Figure~\ref{fig:compare-v18-thiswork-hist}. There is a high degree of agreement between the two radius distributions, especially regarding the location of the radius gap. Subtle differences remain due to residual offsets in host star properties, differences in planet detectability, or other effects. Finally, we emphasize that all distributions shown represent detected planets, not completeness-corrected occurrence rates. We will leave such caveats to future studies.

\section{Future Work}
\label{sec:future-work}

\subsection{Investigating Gap Planets}

The planets that remain in the radius gap after our duration filter are intriguing and warrant further investigation. A number of observational and astrophysical explanations should be explored. Even if the radius gap were completely devoid of planets, we would still expect some spill-over due to statistical scatter because the gap planets reside less than 4$\sigma$ from the edge of the super-Earth/sub-Neptune populations.

Aside from statistical uncertainties, we expect a small number of high-$b$ planets to pass our duration filter due to $e$ effects. When this occurs for a sub-Neptune, our measured radius is too small by 10--20\% and the planet enters the gap from above. Based on our simulations (\S\ref{sec:filter}), we expect that this scenario applies to no more than 3 planets out of 100; however, our sample includes hundreds of sub-Neptunes, so this effect must account for a handful of gap planets in Figure~\ref{fig:compare-per-prad}.

There are other sources of \Rp bias that are unrelated to the photometric modeling discussed in this paper. When the \Kepler team produced photometry, they worked to remove the contribution of stray light in their software apertures \citep{Batalha10}. However, this dilution correction could only account for sources listed in the \Kepler Input Catalog
%
\citep{Brown11}. The transit depths of planets with stellar companions within a few arcsec include uncorrected dilution, which reduces their apparent size. This is another way in which sub-Neptunes could spill into the radius gap. F18 removed as many of these cases as possible using the \cite{Furlan17} compilation of high-resolution imaging. However, such observations do not exist for all \Kepler, and the observations themselves rarely rule out all parameter space where diluting companions could reside. 

We must also consider the possibility that these gap planets actually straddle the two populations and the astrophysical implications of such planets. Current theories to explain the radius gap invoke a core mass distribution that declines sharply between roughly 5 and 10~\Me. In both the star-powered and core-powered mass-loss models, planets gradually lose their envelopes until the envelope fraction drops below $\sim$1\%. Then, envelope-loss accelerates and the planets quickly hop the gap to become envelope-free bodies \citep{Owen17,Gupta19}. Some of the gap planets may be in the process of this transition and present exciting prospects for observing  mass-loss in realtime. 

Another possible formation channel is planet mergers. For example, the merger of two Earth-composition cores of 5~\Me and 1.5~\Re in size would produce a 10~\Me planet with a size of 1.8~\Re. Given the various effects that can move planets into the radius gap, we recommend further observational and photometric modeling efforts to pin down their radii. However, if the radii of these gap planets are correct, they should provide a powerful diagnostic of planet formation models.

\subsection{Refitting Kepler Photometry}

In this paper, we have shown one can leverage short-cadence photometry and knowledge of $e$ and \rhostar to reduce the bias in measurements of \Rp/\Rstar and hence \Rp. To date, there has yet to be a systematic effort to refit \Kepler lightcurves with $e$-\rhostar priors or to include all available short-cadence data. This would certainly require significant effort, beyond the scope of this paper. Here, we take stock of the available datasets, quantify their potential value-add, and sketch a prescription for a re-analysis of \Kepler photometry.

Today, \Kepler host stars are much better characterized compared to the prime mission thanks largely to \Gaia. In particular, \rhostar measurements with 12\% uncertainties are typical. \cite{Van-Eylen19} and others have characterized the eccentricity distribution of \Kepler singles and multis, which can serve as priors during the light-curve modeling. Finally, of the 4078 planets candidates in T18, 1138 have at least one quarter (90 days) of short-cadence photometry. Of these, 437 are in the curated F18 sample. While not complete, this is significantly more than the V18 sample of 117. 

We quantified the potential improvement in \Rp/\Rstar when including this additional information through several representative fits to synthetic photometry. We generated 17 quarters of synthetic short-cadence photometry with the {\em exoplanet} package \citep{exoplanet:exoplanet}. Of the 437 planets discussed above, the median target had $P$ = 11.5~days, \Rp/\Rstar = 1.7\%, and S/N = 48. Our simulated transit had the same $P$ and $\Rp/\Rstar$. We set $\{b, e, \rhostar, u_1, u_2\} = \{0.9, 0.0, 1.0~\mathrm{g/cc}, 0.45, 0.23\}$ and injected Gaussian noise into the photometry to achieve S/N = 50.

We then performed three MCMC analyses of this dataset with varying degrees of binning. We fit light curves with 0, 1, and 17 quarters of short-cadence. In our modeling, we allowed the following parameters to vary: $\{P, T_c, \Rp/\Rstar, \rho_{\star,tr}, b \}$. We sampled the posterior using the No-U-Turn Sampler (NUTS) implemented in the {\em Exoplanet} package \citep{exoplanet:pymc3,exoplanet:exoplanet}. After tuning, we sampled the chains until they were at least 1000$\times$ longer than the auto-correlation length.

\begin{figure*}[h!]
\centering
\includegraphics[width=0.99\textwidth]{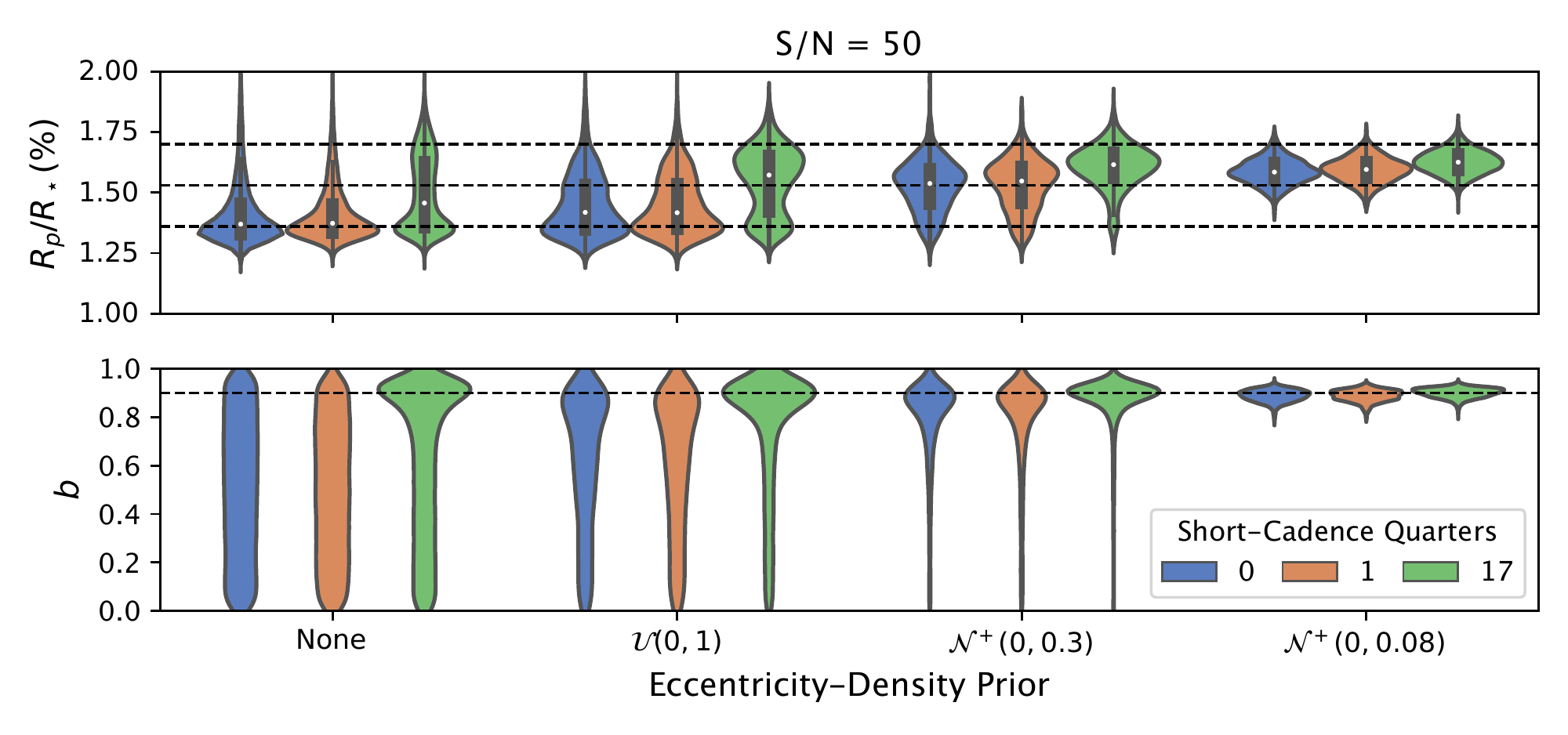}
\vspace{-0.25cm}
\caption{The ``violins'' show the posterior probability density of $\Rp/\Rstar$ (top) and $b$ (bottom) based on fits to synthetic photometric datasets. For the input transit model, $\Rp/\Rstar = 1.7\%$, $b = 0.9$, and S/N = 50. Violins are color-coded by the amount of short-cadence: blue---none, orange---1 quarter (90 days), and green---17 quarters. Violins are grouped according to different $e$-$\rhostar$ priors: The first group has no prior. The other groups include a Gaussian prior on \rhostar with 12\% dispersion and the following priors on $e$: $\mathcal{U}(0,1)$, $\mathcal{N}^+(0,0.3)$, and $\mathcal{N}^+(0,0.08)$. The dashed lines correspond to 1.0, 0.9, and 0.8 times the input $\Rp/\Rstar$. With no $e$-$\rhostar$ prior, the $\Rp/\Rstar$ posterior medians (white dots) are biased by $\approx20\%$ because $b$ is nearly unconstrained. As we impose tighter $e$ priors, the bias on $\Rp/\Rstar$ decreases and $b$ convergences to 0.9. Under the $e \sim \mathcal{N}^+(0,0.08)$ prior, applicable for multi-transiting planets, the bias is 5\%. At this S/N, long- and short-cadence fits yield similar constraints; there is insufficient S/N to measure ingress/egress durations from short-cadence to constrain $b$.}
\label{fig:violin-snr-50}
\end{figure*}

Figure~\ref{fig:violin-snr-50} shows the constraints on $\Rp/\Rstar$ and $b$ from these fits. For all binnings, with no $e$-$\rhostar$ prior, $b$ is nearly unconstrained and the median of the \Rp/\Rstar posterior is biased below its true value by 15-20\%. We then imposed different $e$-$\rhostar$ priors by re-weighting the posterior samples using importance sampling. We imposed a 12\% Gaussian prior on $\rhostar$. Applying a uniform eccentricity prior, encouraged the $b$ posterior probability to concentrate around the input value $b = 0.9$. For the short-cadence only dataset, this was sufficient to reduce the bias of the median $\Rp/\Rstar$ to $<10\%$; however, there was little change to the binnings with zero or one quarter of short-cadence.

We also applied a prior on $e \sim \mathcal{N}^{+}(0,0.3)$, which is justified for \Kepler singles \citep{Van-Eylen19}. This further concentrated the $b$ posterior around 0.9 and brought the median $\Rp/\Rstar$ to within 10\% of its input value for all binnings. Applying a still tighter $e \sim \mathcal{N}^{+}(0,0.08)$ prior, which is justified for \Kepler multis, further reduced the bias to $\approx$5\% for all binnings.

These experiments illustrate the relative value of short-cadence and $e$-\rhostar priors. While there are many planets with one or more quarter of short-cadence, at the median S/N, short-cadence alone is only marginally useful in identifying the high-$b$ transits that result in the largest $\Rp/\Rstar$ biases. However, physically motivated $e$-$\rhostar$ priors can significantly reduce these biases, even when only long-cadence is available.

\begin{figure*}[h!]
\centering
\includegraphics[width=0.99\textwidth]{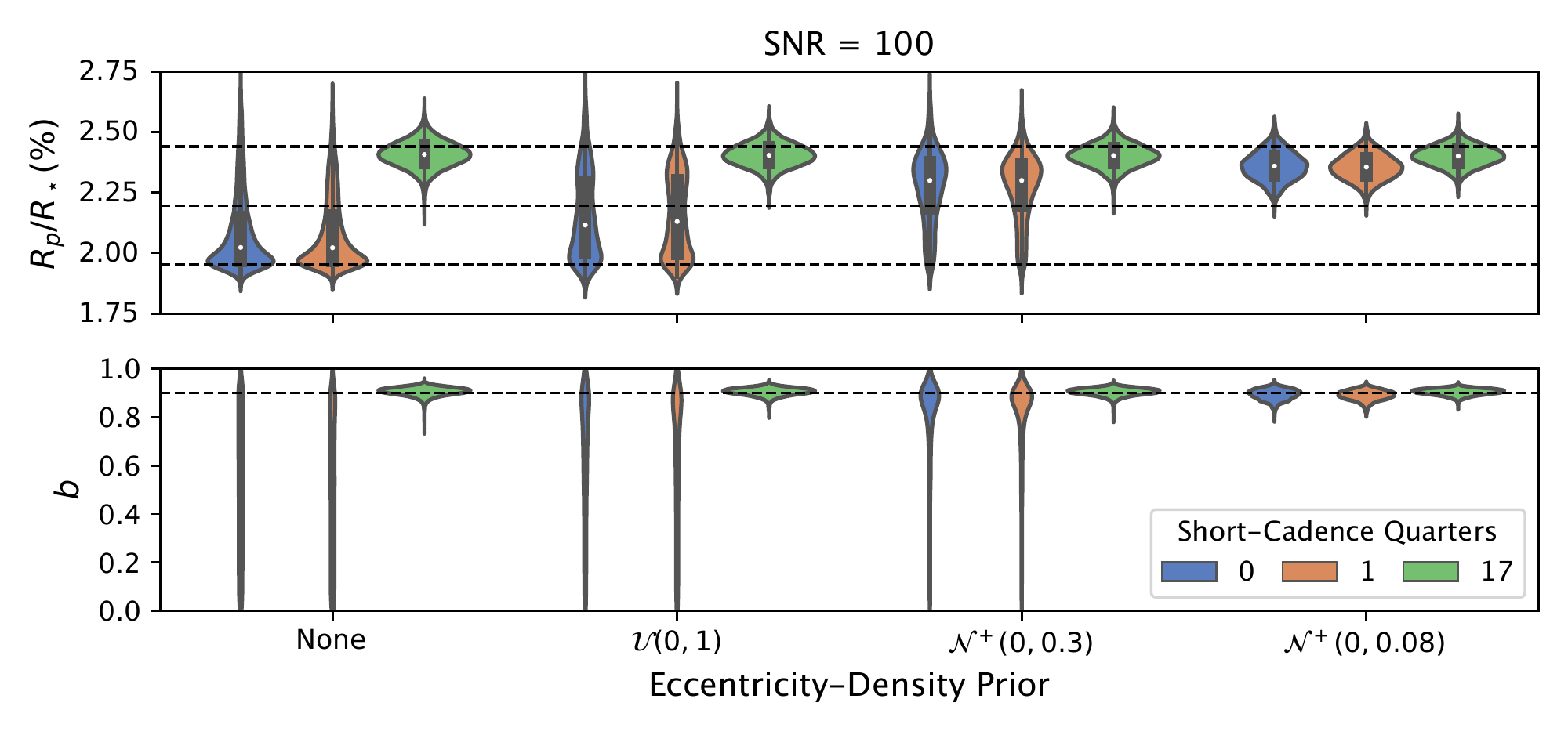}
\vspace{-0.25cm}
\caption{Same as Figure~\ref{fig:violin-snr-50} except input transit has \Rp/\Rstar = 2.45\% and S/N = 100. The larger \Rp/\Rstar and S/N permits a tight constraint on $b$ from short-cadence fits (green), which results in an unbiased measurement of $\Rp/\Rstar$ with no $e$-$\rhostar$ prior. For all sampling rates, adding physically motivated $e$ priors, $\mathcal{N}^+(0,0.3)$ or $\mathcal{N}^+(0,0.08)$, reduces the bias on $\Rp/\Rstar$ to $\approx$5\%. There is very little difference in the posterior medians between the datasets with zero and 90 days of short-cadence data. At this S/N, one cannot reliably constrain $b$ from 90 days of short-cadence alone.}
\label{fig:violin-snr-100}
\end{figure*}

At higher S/N, it is possible to measure ingress/egress durations with sufficient precision to constrain $b$ from short-cadence alone. Figure~\ref{fig:violin-snr-100} is identical to Figure~\ref{fig:violin-snr-50}, except that the simulated transit had \Rp/\Rstar = 2.45\% and S/N = 100. As before, the fits with zero or one quarters of short-cadence could not constrain $b$ and $\Rp/\Rstar$ was biased by $\approx$20\%. However, the fits to only short-cadence data converged on the input $b$ and $\Rp/\Rstar$ within statistical uncertainties.

These simulations motivate a re-analysis of \Kepler light curves using $e$-$\rhostar$ priors and short-cadence data, when available. Of course, such priors on $e$ are based upon previous studies of \Kepler light curves. The most self-consistent approach would be to simultaneously model the properties of the individual planets ($\Rp/\Rstar$, $b$, etc), along with the population-level $e$ distribution. Bayesian hierarchical modeling provides a framework for such an analysis. 

\subsection{Other instruments}

Of course, new photometry could also be helpful. {\em TESS}  and  {\em PLATO} may observe large numbers of \Kepler transits at 2~min and 25~s sampling, respectively \citep{Ricker14,Rauer13}. However, larger photon-limited uncertainties due to the smaller effective apertures of these telescopes may limit their ability to refine \Rp. For high-value planets, such as those falling in the radius gap, targeted observations by {\em CHEOPS} \citep{Cessa17CHEOPS} and {\em JWST} may improve \Rp. Observations in infrared would be especially valuable as less limb-darkening diminishes the covariance between $b$ and $\Rp/\Rstar$.

Finally, multi-color photometry provides additional leverage on impact parameter due to the wavelength dependence of limb-darkening (see, e.g., \citealt{Tingley04}). In principle, multi-band observations by instruments such as MuSCAT \citep{Narita19} could break some of the $b$--\Rp/\Rstar degeneracies discussed here, but lower photometric precision may make such work impractical from the ground. We recommend that future space-based transit projects consider the potential improvements to planet radii as an additional value-add of multi-wavelength capabilities.

\section{Conclusions}
\label{sec:conclusions}

In this paper, we scrutinized differences between the F18 and V18 views of the radius gap. We showed that different \Rstar cannot account for the difference between these two studies. Comparing the \Rp/\Rstar derived by different techniques illustrated some of the challenges that arise when performing population statistics with the NEA tables, which includes noise in \Rp/\Rstar from the fitting (\S\ref{sec:comparison}). We made two recommendations for precision studies of \Kepler planet radii: (1) adopt the posterior medians and (2) discard high-$b$ planets. While it is usually not possible measure $b$ with sufficient precision from long-cadence photometry to perform (2), transit duration can serve as an effective proxy (\S\ref{sec:filter}). We applied these practices to the CKS dataset and increased the contrast between the super-Earth and sub-Neptune populations. We found that differences in \Rp/\Rstar and the distribution of host star masses both contribute to the differences between (\S\ref{sec:cks}). Finally, we offered some suggestions for future studies of the \Kepler population (\S\ref{sec:future-work}).

The discovery of abundant planets between the size of Earth and Neptune is one of the most profound results from the \Kepler mission. That these planets come in two size classes is a powerful signpost of their formation channels. Theoretical work to understand the super-Earth and sub-Neptune populations is ongoing. We can only wonder what additional features will emerge as we characterize these planets to still higher precision.

\acknowledgments
We thank Josh Winn, Vincent Van Eylen, and Dan Foreman-Mackey for constructive comments that improved this manuscript. We are grateful to the anonymous referee for their careful review and suggestions. E.A.P. acknowledges support from the NASA Astrophysics Data Analysis Program (ADAP) grant 80NSSC20K0457.

\software{{\em Astropy} \citep{exoplanet:astropy18}, {\em Exoplanet} \citep{exoplanet:exoplanet}, {\em pandas} \citep{pandas}, {\em PyMC3} \citep{exoplanet:pymc3}, {\em numpy/scipy} \citep{numpy/scipy}, {\em STARRY} \citep{exoplanet:luger18}, {\em Theano} \citep{exoplanet:theano}, and \cite{exoplanet:agol19}.}

\bibliography{manuscript_v2}
\end{document}